\documentstyle[preprint,aps,graphicx]{revtex}
\begin{document}
\draft
\title{ Is The Heavy Fermion Formalism Applicable to Meson Production 
Processes ?} 
\author{ E. Gedalin\thanks{gedal@bgumail.bgu.ac.il},
 A. Moalem\thanks{moalem@bgumail.bgu.ac.il}
 and L. Razdolskaya\thanks{ljuba@bgumail.bgu.ac.il}}
%\date
\address{ Department of Physics, Ben Gurion University, 84105
Beer Sheva, Israel}
\maketitle
\begin{abstract}
We consider tree diagram contributions to neutral pion 
production in $pp \rightarrow pp \pi^0$ using
chiral perturbation theory in the relativistic and extremely
non-relativistic limits. In marked difference with the results 
from heavy fermion formalism, the impulse and s-wave rescattering terms 
in the relativistic limit have equal signs and therefore 
add constructively, giving rise to a substantial contribution to
the cross section. We argue that the 
power series expansion of the nucleon propagator is on the border
of its convergence circle. Consequently, a $finite ~order$ heavy fermion 
formalism  does not predict nucleon pole terms correctly and therefore 
can not be applied to meson production.\\

\end{abstract}


\newpage

In recent years, intensive theoretical efforts have been devoted to 
investigating  how nuclear and hadron interactions can be 
understood within Chiral Perturbation Theory ($\chi$PT), 
an approach which is generally believed to be an effective 
theory of Quantum Chromo Dynamics (QCD) at low energies.  This is an 
important clue toward understanding nuclear dynamics within the 
context of QCD, the accepted fundamental theory of strong interactions. 

 An outstanding, and still unsolved, problem in this area 
arises in the understanding of neutral pion production in 
$pp \rightarrow pp \pi^0$. Park et al.\cite{park96}, Cohen et
al.\cite{cohen96} and Sato et al.\cite{sato97} 
have considered this process in $\chi$PT by applying rather 
similar calculation schemes based on extremely non-relativistic 
heavy fermion formalism (HFF), where the leading order impulse 
(graphs 1a-1b) and rescattering (graph 1c) contributions are 
found to have opposite signs, and hence leading to a cross section
substantially smaller than experiment\cite{park96,cohen96,sato97}. 
In this regard, it is to be noted that meson-exchange 
models predict equal signs for the two terms, achieving quite
impressive descriptions of data near threshold
\cite{hernandez95,hanhart95,gedalin96,hanhart98,tamura98}.
 Particularly, in covariant
exchange models\cite{hernandez95,gedalin98} the amplitudes from the
impulse and rescattering terms interfere constructively.  
It has been argued by Hanhart et al. \cite{hanhart98} that this 
sign difference between predictions of meson exchange models and
HFF $\chi$PT is a genuine feature. More recently, Tamura et 
al. \cite{tamura98} have concluded that the shape of 
energy spectra for the $d(p,(pp)_s)\pi^0 n$  and
$d(p,(pp)_s)\pi^- p$ reactions, can be explained  
only if the interference between these terms is constructive.

By extending calculations to chiral order $\Delta =2$ Gedalin et
al.\cite{gedalin98} have shown that in addition to not predicting
correctly the  relative phase of the leading order impulse 
and rescattering terms, the HFF yields one loop contributions 
substantially larger than lower order terms. This is a serious drawback
which indicates that the HFF expansion converges (if at all) slowly 
and therefore may not be suitable to apply to production processes. 
It is the purpose of the present note to call attention to the 
fact that in a relativistic $\chi$PT the sign and relative importance
of various contributions are different from those found using  HFF.
First, the contribution from rescattering is substantially larger 
and having an equal sign as the impulse term. Secondly, it is to 
be shown that these differences are mostly due to the fact that 
in the reduction procedure of the pion-nucleon Lagrangian, the
nucleon kinetic term is reduced also, what limits the
validity of the HFF Lagrangian to sufficiently small nucleon momenta.
Consequently, $a ~finite ~order$ HFF can not be applied to meson
production 
processes which necessarily involve large momentum transfers.

To begin with we consider the tree diagram contributions  (Fig. 1) 
to neutral pion production in $pp  \to pp \pi ^0 $.  
Let $\pi$ and $N$ represent the pion and nucleon fields, 
$m$ and $M$ the pion and nucleon masses,  then the 
fully relativistic pion-nucleon sector 
Lagrangian\cite{gasser85,gasser88,park93} is,  
\begin{eqnarray}
& & L_{\pi N} = {\bar N} (D\! \! \! \! /^{(1)} - M) N + 
\frac {c_1' m^2}{2M} {\bar N} \langle U^{\dagger} + U \rangle N + 
\frac {c_2'}{4 M^3} {\bar N} \tensor {D}^{\mu} \tensor {D}^{\nu}
N\langle \Delta_{\mu} \Delta_{\nu} \rangle - \nonumber \\
& & \frac {c_3'}{M} {\bar N} N \langle \Delta \cdot 
\Delta \rangle + 
\frac {c_4'}{2M^2} {\bar N} i \gamma^{\mu} \tensor {D}^{\nu} N
\langle \Delta_{\mu} \Delta_{\nu} \rangle~.
\label{lagrang}
\end{eqnarray}
Here $F = 93$ MeV and $g_A = 1.26$ are the pion radiative 
decay and axial vector coupling constants; and $c_i ' (i = 1-4)$ 
are the low energy constants\cite{park93,bernard95,bernard97,fettes98}.
The covariant derivatives $D_{\mu}$, $\Gamma_{\mu}$ and
axial vector field $\Delta{\mu}$ are related through the expressions :
\begin{eqnarray}
& & D\! \! \! \! / ^{(1)} = D\! \! \! \! / + i g_A {\Delta}\! \! \! /
\gamma ^5~,
\\
& & D_{\mu} = \partial_{\mu} + \Gamma_{\mu}~, \\
& & \Gamma_{\mu} = \frac {1}{2} 
\left[ \xi^{\dagger}, \partial_{\mu} \xi \right]~,\\
& & \Delta_{\mu} = \frac{1}{2}
\left\{\xi^{\dagger}, \partial_{\mu} \xi \right\}~.
\end{eqnarray}
Here $D\! \! \! \! / = \gamma ^{\mu} D_{\mu}$;
$ {\bar N} \tensor {D} N  =
{\bar N}(DN)  -  (D^{\dagger}{\bar N}) N $;
the symbol   $\langle B  \rangle$ stands for the trace of the 
quantity B over isospin matrices and, 
a nonlinear realization of the chiral symmetry  $U = \xi ^2 = 
\exp(i{\bf \tau}\cdot {\bf \pi}/F)$ is presumed to describe 
the pion field. 

The pion-nucleon scattering amplitude within the framework of a 
relativistic $\chi$PT has been derived already by Gasser et al.
\cite{gasser88}. Using their results, the  Lagrangian Eqn.
\ref{lagrang}, and standard Feynman rules, the nucleon pole 
and sea-gull (diagrams 1a-1c) contributions to the production
process are,
\begin{eqnarray}
&{\cal M}^{(1)}_I = N_1~ N_3~ \frac{ig_A}{2F (q^2 - m^2)}~
T^{(s+u)}~
\chi_3^{\dagger}~({\bf \Pi}_1 - {\bf \Pi}_3)\cdot{\bf \sigma }_1~\chi_1
+ [1 \leftrightarrow 2, 3 \leftrightarrow 4]~,\nonumber \\
\label{impulse}
&{\cal M}^{(1)}_R = N_1~ N_3~ \frac{i g_A}{ 2F (q^2 - m^2)}~
T^{(c)}~
\chi_3^{\dagger}~({\bf \Pi}_1 - {\bf \Pi}_3)\cdot{\bf \sigma }_1~\chi_1
+ [1 \leftrightarrow 2,3 \leftrightarrow 4]~,\nonumber\\
 \label{rescat}
\end{eqnarray}
with,
\begin{eqnarray}
& & T^{(s+u)} = N_2 N_4 \frac {g_A^2 M^2}{F^2} \chi_4^{\dagger}
\left[ \frac {1}{ (p_4+k)^2 - M^2} k {\it P} - 
\frac {1}{ (p_2-k)^2 - M^2} q {\it P} -
\frac {1}{M} {\it R} \right] \chi_2 ~,
\label{supoles}\\
& & T^{(c)} = - N_2 N_4 \frac {m^2}{MF^2} \chi_4^{\dagger}\nonumber \\
& & \left\{-2c_1' + c_2' \frac {q(p_4 + p_2) k(p_4 + p_2)}
{4m^2M^2} + c_3' \frac {kq}{m^2} + c_4' \frac {1}{4M m^2} 
\left[ (p_4 + p_2)q k {\it P} + 
(p_4 + p_2)k q{\it P}\right] \right\} \chi_2~.
\label{seagull}
\end{eqnarray}
Here,
\begin{eqnarray}
& & N_i = \sqrt {\frac {E_i + M} { 2M} }~,~~~~
{\bf \Pi}_i =\frac {{\bf p}_i}{E_i + M}~,\nonumber \\   
& & {\it P} = \left( 1 + {{\bf \Pi}_2} \cdot {{\bf \Pi}_4} +
i {{\bf \Pi}_2} \times {{\bf \Pi}_4} \cdot {\bf \sigma}_2~,
{\bf \Pi}_2 + {\bf \Pi}_4 + i ({\bf \Pi}_2 - {\bf \Pi}_4) \times
{\bf \sigma}_2\right)~,\nonumber \\
& & {\it R} = 1 - {\bf \Pi}_2 \cdot {\bf \Pi}_4 +
i {\bf \Pi}_2 \times {\bf \Pi}_4 \cdot {\bf \sigma}_2~,
\end{eqnarray}
where  
$q =  p_1 - p_3,~ k = (\sqrt {m^2 + {\bf k}^2}, {\bf k}),
~p_1 = (E({\bf p}), {\bf p}),~ p_2 = (E({\bf p}), -{\bf p}),
~p_3 = (E({\bf p'}-{\bf k}/2), {\bf p'}-{\bf k}/2),
~p_4 = (E(-{\bf p'}-{\bf k}/2), -{\bf p'}-{\bf k}/2)$ 
stand for the momenta of the incoming and outgoing pions 
and nucleons (see Fig. 1) in the overall
center of mass (CM) system, respectively.
The bracket $[1 \leftrightarrow 2, 3 \leftrightarrow 4]$
represents the contribution from the same diagrams with
the proton momenta $p_1,\ p_3$ interchanged with $p_2,\ p_4$,
respectively.
Note that $T^{(s+u)}$ and $T^{(c)}$ are analogous impulse and 
sea-gull contributions (graphs 2a-2c) to the 
$\pi^0 N \to \pi^0 N$ conversion process, as derived by
Gasser et al.\cite{gasser88}. 
 
At low scattering energy and near threshold of the 
$pp \rightarrow pp \pi^0$ reaction, 
${\it P} = (1,{\bf \Pi}_2 + i{\bf \Pi}_2 \times {\bf \sigma}_2), 
{\it R} = 1$ so that,
\begin{equation}
T^{(s +u)} \approx \frac {g_A^2 M^2}{F^2} N_4 N_2 
\frac {\it T}{M (2 M k^0 + k^2 )(2Mq^0 - q^2)}~,
\end{equation}
with,
\begin{equation}
{\it T} \approx k^4 - {\bf q}\cdot {\bf q} k^0 
\left(k^0 + \frac {k^2}{M}\right)~.
\end{equation}
The quantity ${\it T}$ determines the magnitude as well as
the sign of $T^{(s+u)}$. When both of the incoming and outgoing 
pions are on the mass shell( i.e. $k^2 = q^2 = m^2$ and 
${\bf k}^2, {\bf q}^2 \ll m^2$) it is positive, ${\it T} = m^4 > 0$. 
For the production process however, only the outgoing pion is on 
the mass shell. The exchanged pion is far off mass shell ($q^2 \leq -
M m , {\bf q}^2 \geq M m$) and ${\it T}$ is negative, i.e. 
\begin{equation}
T_{off}^{(s+u)} = -0.9/m~.
\end{equation}
We recall that in variance with this value the overall s and u
nucleon pole contribution to lowest order in HFF is always
positive\cite{cohen96},
\begin{equation}
T_{HFF}^{(s+u)} = \frac {g_A^2}{4MF^2}\left({\bf q}^2 + {\bf k}^2\right)~.
\end{equation}
Thus in the transition from a fully relativistic theory to an extreme 
non-relativistic limit, the contribution from nucleon pole terms to 
off mass shell $\pi$N scattering amplitude reverses sign. As we shall
demonstrate below, this is mostly because the HFF power expansion of
the nucleon propagator can not be approximated by any finite sum. 

Next consider the sea-gull contribution. 
For low energy scattering and when both pion legs are on 
the mass shell,  Eqn. ~\ref{seagull} reduces to exactly the
expression from HFF, i.e.
\begin{equation}
T^{(c)} \approx - \frac {m^2}{MF^2 } N_2 N_4
\left[ -2c_1' + c_3' 
\left( 1+ \frac { {\bf k}^2 - {\bf q \cdot k}} {m^2}\right) +\left(c_2' +
c_4'\right)
\left( 1 + \frac {{\bf k}^2}{m^2}\right)\right]~. 
\end{equation}
However, off the mass shell and at threshold of the production process, 
Eqn. ~\ref{seagull} becomes,
\begin{equation}
T_{off}^{(c)} = - \frac {m^2}{MF^2} \sqrt {1 + \frac {m}{4M}} 
\left[-2c_1' + c_2' \left(1 + \frac { m }{ 4 M }\right)^2 +
c_3' \frac {1}{2} + c_4' \left( 1 + \frac {m}{4M} \right)\right] ~.
\label{sgulloff}
\end{equation}
This expression departs strongly from the HFF result\cite{park96,cohen96}
\begin{equation}
T_{HFF}^{(c)} = - \frac {m^2}{MF^2}
\left[-2c_1' + \frac {c_2' + c_3' +  c_4'}{2} \right] ~,
\end{equation}
where a term of the order of $(c_2'+c_4')1/2$ is missing.
Clearly, off the mass shell, the relativitistic expression for the sea-gull
contribution ( Eqn. \ref{seagull}) decreases far more rapidly with
the momentum transferred $q$ as compared with 
the respective HFF expression (Eqn. ~\ref{sgulloff}). 

Here as well, this difference between the two approaches
is the result of extending the application of the HFF Lagranagian,
which is limited to low mometum transfer, outside its domain of validity.
By doing so, a $q$ momentum dependent part of the sea-gull term which
is actually of the same order of magnitude, becomes part of higher 
chiral order terms and thus introducing sizeble discrepancy.

In table I we list values of $T^{(c)}$ as obtained with different LEC
parameter sets. The parameter Set 1 is obtained from a tree level
fit of $\pi$N scattering data; the Set 2,3 are proposed in Ref.
\cite{sato97} by extracting the value of $c_2' +c_3'+c_4'$ from the
effective
range parameter $b^+$ of the low-energy pion-nucleon scattering amplitude;
The Set 4 is determined by fitting pion-nucleon scattering to one loop
order\cite{bernard97}; The Sets 5-7 are determined in Ref.
\cite{fettes98} from fitting data to chiral order $\Delta = 2$.  
The magnitude of the sea-gull term is small on the mass 
shell but its sign depends on the values of the LEC used; 
$T^{(c)}$ is negative
with  parameter Sets 2, 6 and 7 but positive with the others. 
Off the mass shell the sea-gull term is rather sensitive to
different choices of the LEC parameter sets, though its sing is 
negative always. A fully relativistic approach predicts  a factor 2-3
larger a contribution as compared to that from HFF. Consequently,
in a fully relativistic approach the nucleon pole terms and the sea-gull
term add constructively. Their overall contribution, which appears as
$T_{off}^{full}$ in Table 1, is strongly enhanced
in comparison with the respective quantity $T_{offHFF}^{full}$ from HFF.
This has a dramatic impact on the calculated cross section of 
the $pp \rightarrow pp \pi^0$ reaction.

In Fig. \ref{figpirel0} we draw cross sections calculated, taking into 
account contributions from the impulse and rescattering terms only.
Corrections due to initial state interactions (ISI) and final state
interactions (FSI) 
are introduced using the approximation II of
Ref.
\cite{gedalin98}. The cross sections calculted with the diffrent LEC 
sets vary within a factor of 2 and, except for results obtained with
the  Set 2, all others
underestimate data\cite{bondar95,meyer92} by a factor 1.5-2. 
We would like to stress though that the curves presented in 
Fig. ~\ref{figpirel0} do not account for contact  terms 
and loop contributions, and therefore may serve for illustrative
purposes only, not for explaining data. The inclusion of the 
contact terms (Fig. 1d-1f) 
must awaits a reliable ( and consistent
with NN interactions) determination of many
LEC's\cite{bernard95,ordonez96}. Unfortunately,
it is not straightforward to extend the procedure previously 
applied in the context of the HFF
\cite{park93,cohen96,kolck96,gedalin98} to determine the
parameters for the contact terms (Fig. 1d-1f)
\cite{ordonez96,gedalin981}.

It is to be noted also, that we have not included lower order 
loop contributions. Restricting our discussion to one loop
only, there are more than 40 possible diagrams which may contribute
to the production process. In the HFF $\chi$PT
calculations\cite{gedalin98}, many of these are
proportional to three momentum of the meson produced and therefore
their contribution to the cross section at threshold is negligibly
small. In the relativistic $\chi$PT these contributions become
proportional to the pion mass and can no longer be neglected. 
Because of fine mutual cancellations amongst them,  all these
diagrams must be included in a systematic way as in Ref. 
~\cite{gedalin98}. However, it should be noted that an inspection of the
expressions derived by Gasser et al. \cite{gasser88} for analogous
contributions to  
$\pi^0 N \to \pi^0 N$, reveals  that such one loop contributions 
to the production process are smaller than the overall 
contribution from the impulse and rescattering  terms.

Clearly, the relative importance of various contributions and 
their signs in a fully relativistic $\chi$PT approach differ
significantly from those obtained in the non-relativistic HFF approach.
To point out the origin of these differences we recall that in
the transition from  a relativistic  $L_{\pi N}$
to non-relativistic HFF Lagrangian one reduces 
the nucleon kinetic term which affects the nucleon propagator
strongly. By doing so, the validity of the HFF Lagrangian  
becomes limited to sufficiently small momenta and therefore is 
not applicable to meson production.
Both of the  s and u pole diagrams, Fig. 1a, 1b, involve a covariant
nucleon propagator,
\begin{equation}
S_N = i \frac {p\! \! / + M}{p^2 - M^2}~,
\label{propag}
\end{equation}
which we may separate into a positive and negative energy
parts. To do this let us write the numerator in Eqn. \ref{propag} as
\begin{eqnarray}
& & p\! \! / + M = M (1 + v\! \! \! /) + (p\! \! / - Mv\! \! \! /)
= 2M P_+ + l\! \! /(P_+ + P_-)~,
\end{eqnarray}
where $v^{\mu}$ denotes the four velocity of the nucleon with $v^2 = 1$,
$l^{\mu} = p^{\mu} -M v^{\mu}~$ is its residual momentum, 
and $P_{\pm} = (1/2)(1 \pm v\! \! \! /)$ are operators which 
project the nucleon Dirac field into large and small components.
Following the usual reduction procedure, we take $p_{\bot} = p^\mu -
(pv)v^\mu$ to be the
transversal component of the nucleon momentum and
write the nucleon propagator in the form,
\begin{equation}   
S_N = i \left[2M P_+ + l\! \! \! / (P_+ +P_-)\right]
\frac {1}{2 (M + T(p_{\bot}))}\left[ \frac {1}{vl - T(p_{\bot})} -
\frac {1} {2M +vl + T(p_{\bot})}\right]~.
\label{propagator}
\end{equation}
with $T(p_{\bot}) = \sqrt {M^2 + p_{\bot}^2} -M$.
In the limit of
low kinetic energy the negative energy part reduces to
$S_N \approx -  {i P_+}/{2M}$. It is easy to show now that, 
the respective negative energy contributions
from s and u channels sum up to be,
\begin{equation}
T_Z \approx \frac {g_A^2}{4MF^2} (vq)(vk)~.
\label{contact}
\end{equation}
This in fact has the form of a contact term. In passing by, we note
that by separating
the nucleon propagator into negative and positive energy parts   
the s and u channels (graphs 1a, 1b) split into direct and 
Z-graph contributions\cite{alfaro}. In the non-relativistic limit 
the Z-graph contribution appears in exactly the form of Eqn.
\ref{contact}, as a local rescattering term. Thus in the transition
to non-relativistic limit, the negative energy part of the nucleon 
pole terms "converts" into a sea-gull contact term.
An even more serious a drawback concerns the direct part of 
the nucleon pole terms. It is not always possible to calculate 
the direct part, which is a non-local term, within the frame of HFF.
To see this consider the expansion of
$S_N$, Eqn. \ref{propagator} in power series.
The series expansion for the factors $1/(M + T(p_{\bot})$ and 
$1/(2M + vl + T(p_{\bot}))$
converges up to high energies. However, the series
\begin{equation}
\frac {1}{vl - T(p_{\bot})} = \frac {1}{vl}
\sum \left[ \frac {T(p_{\bot})}{vl}\right]^n~,
\end{equation}
converges only for  $T(p_{\bot}) / vl < 1$. 
In the instance of $\pi N \to \pi N$
scattering and in the limit $T(p_{\bot}) \ll vl$ this series 
converges rather well. However for a production process
$NN \rightarrow NN X$, the virtual nucleon in the graph 1b has  
a residual momentum $l = (-m_X/2, {\bf l}) ; {\bf l} \cdot {\bf l} = M
m_X$,
so that $T(p_{\bot}) / vl \approx -1$. Thus the power series is on
the border of its convergence circle and
therefore it can not be approximated by any finite sum. This 
shows, in fact,  that the HFF can not possibly predict the
impulse term correctly, and therefore excludes
the possibility that a finite chiral order HFF based $\chi$PT 
calculations can explain meson production in NN collisions.

In summary, we have considered tree diagram contributions to 
neutral pion  production in $pp \to pp \pi^0$ in a fully relativistic
$\chi$PT and in the extremely non-relativistic HFF. We have  
found that in the relativistic approach, the relative phase 
of the impulse and rescattering terms is +1 and the two terms add
constructively, giving rise to a substantial contribution to the 
cross section.  This stands in marked difference with the HFF
results where these two terms have opposite signs,      
the rescattering term being considerably smaller and their overall 
contribution to the cross section is small. 
The usefulness and success of non-renormalizable (and renormalizable)
effective field theories, depend on how fast the respective perturbative
expansion converges. It was demonstrated that for meson production in
NN collisions, the HFF series of the nucleon propagator is on the border
of its convergence circle. Consequently, meson production falls
outside the HFF validity domain, making the predictions from a
non-relativistic $\chi$PT for such processes impossible.
Since the preparation of this note, we have notived that the 
J${\ddot u}$lich group\cite{meisner98} have reported on similar results.

\vspace{1.5 cm}
{\bf Acknowledgments} This work was supported in part 
by the Israel Ministry Of Absorption.
We are indebted to  Z. Melamed
for assistance in computation.

\begin{table}
\begin{tabular}{|c|c|c|c|c|c|c|c|c|}
    \hline
	 Set No. & $c'_1$ & $c'_2+c'_4$ &$ c'_3$ & $mT^c_{on}$ &
$mT^c_{off}$ & 
	  $mT^c_{offHFF}$ & $mT^{full}_{off}$ & $mT^{full}_{offHFF}$  \\
	\hline
	1 & -1.63 & 6.28 & -9.86 & 0.135 & -1.53 & -0.48 & -2.41 & 0.42  \\
	\hline
	2 & -1.63 & 8.46 & -9.86 & -0.63 & -2.29 & -0.87 & -3.17 & 0.03  \\
	\hline
	3 & -1.87 & 6.28 & -9.86 & 0.007 & -1.69 & -0.64 & -2.57 & 0.26  \\
	\hline
	4 & -1.74 & 6.28 & -9.94 & 0.08 & -1.59 & -0.54 & -2.47 & 0.34  \\
	\hline
     5 & -2.38 & 6.07 & -11.14 & 0.105 & -1.77 & -0.76 & -2.65 & 0.14  \\
	\hline
	6 & -2.76 & 6.03 & -11.27 & -0.1 & -2.0 & -0.98 & -2.88 & -0.08  \\
	\hline
	7 & -2.87 & 6.05 & -11.63 & -0.054 & -2.0 & -1.0 & -2.88 & -0.1  \\
	\hline
\end{tabular}
\vspace{0.5cm}
	\caption{The sea-gull term and pion-nucleon scattering
amplitude for different LEC parameter sets. $T^c_{on}$ represents
the on mass shell sea-gull term contribution to the pion-nucleon
scattering amplitude.The quantities $T^c_{off}$ and 
$T^c_{offHFF}$ denote the off mass shell sea-gull term in the 
relativistic and HFF limits,respectively. The amplitudes 
$T^{full}_{off}$ and $T^{full}_{offHFF}$ represents the 
respective overall contribution of the impulse and sea-gull 
terms in relativistic and non-relativistic limits.
The pole term is equal to $mT^{s+u}=-0.88$ in the relativistic 
approach and $mT^{s+u}_{HFF}=0.88$ in HFF $\chi$PT. 
The different LEC parameter sets are listed in columns 2-4; the Set 1 
is taken from Ref.\protect\cite{bernard95}, Set 2, 3 from Ref.
\protect\cite{sato97},
Set 4 from Ref. \protect\cite{bernard97} and the Sets 5-7 from 
Ref.\protect\cite{fettes98}.
}
\label{table1}
\end{table}
%\newpage
\begin{table}
\begin{tabular}{|c|c|c|c|c|c|c|c|}
	\hline
	 & 1 & 2 & 3 & 4 & 5 & 6  & 7  \\
	\hline
   	${\cal M\/}$(fm) & 92 & 121 & 98 & 94 & 101 & 110 & 111  \\
	\hline
	${\cal M\/}_{HFF}$(fm) & -16 & -1.1 & -10 & -13 & -5.3 & 3 & 3.8  \\
	\hline
\end{tabular}
\vspace{0.5cm}
	\caption{The full primary production amplitudes ${\cal M\/}$ and 
	${\cal M\/}_{HFF}$ at threshold for different sets of $c'_i$}
\end{table}

%\end{document}

\newpage
\begin{figure}
\includegraphics[scale=0.6]{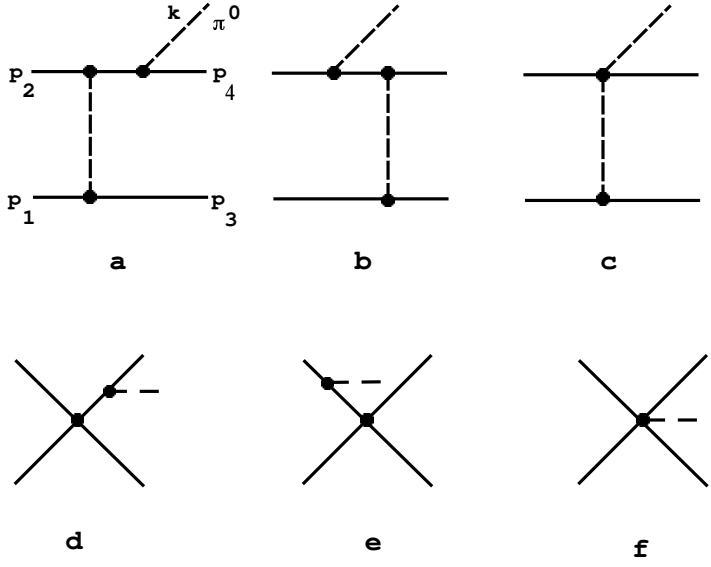} 
\caption{Various tree level diagram to the $NN \to NN \pi^0$ 
reaction : (a) s-channel nucleon pole impulse term, 
(b) u-channel nucleon pole impulse term, (c) rescattering term,
(d)-(f) various contact term contributions. 
}
\label{xfig3}
\end{figure}

\newpage
\begin{figure}
\includegraphics[scale=0.6]{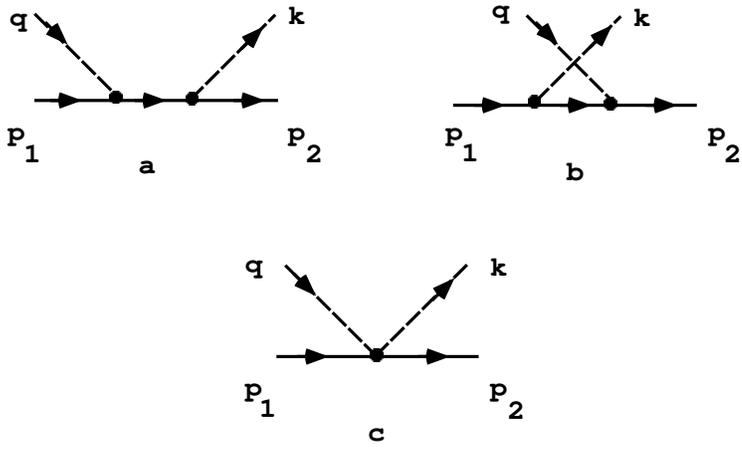}  
\caption{ 
Pole  and sea-gull terms contributing to the
$\pi N \to \pi N$
}
\label{xfig33}
\end{figure}

\newpage
\begin{figure}
\includegraphics[scale=0.80]{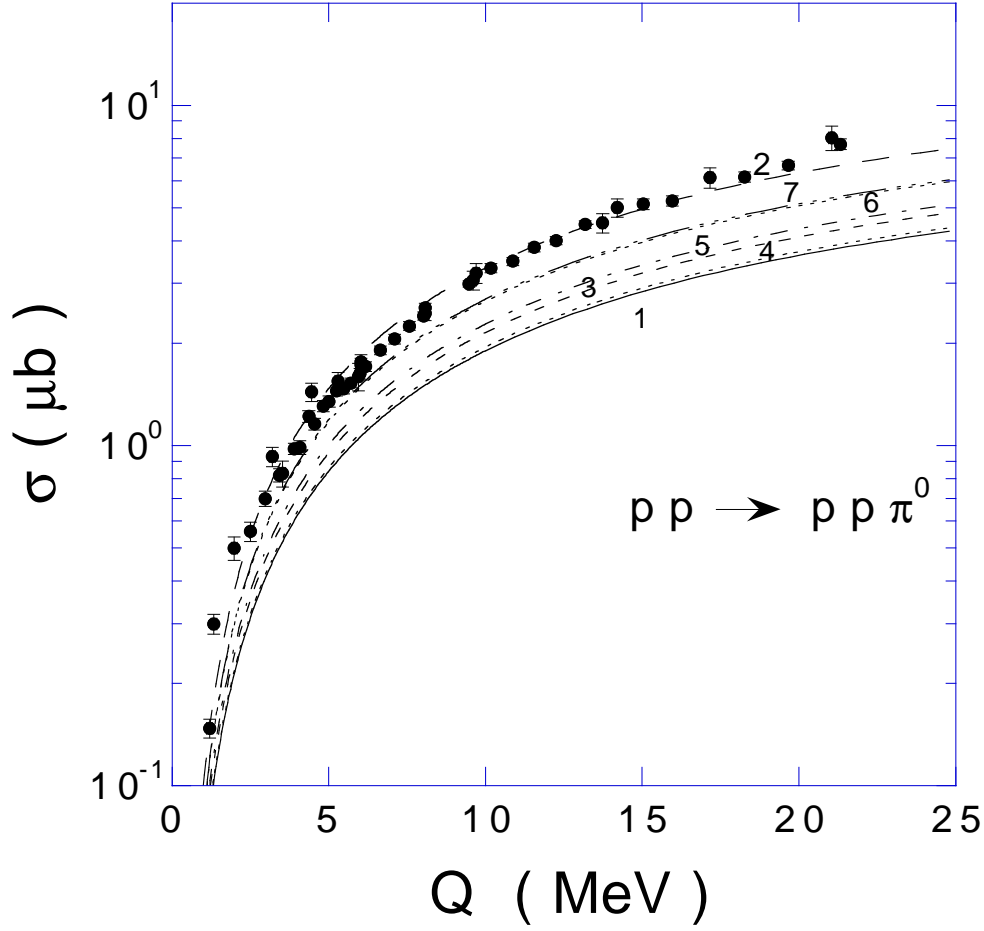}
\caption{S-wave production
cross section for the $pp \to pp \pi^0$ from relativistic $\chi$PT.
Only the impulse and s-wave rescattering terms are included. 
All curves are corrected for ISI and FSI using the approximation II 
of \protect \cite{gedalin98}.
The labels of the curves denote the LEC parameter sets. 
See caption of table \protect
\ref{table1}.
The data are taken from Ref.\protect \cite{bondar95,meyer92}
}
\label{figpirel0}
\end{figure}

\end{document}